\documentclass[10pt, conference, letterpaper]{IEEEtran}
\usepackage[T1]{fontenc}
\IEEEoverridecommandlockouts
\UseRawInputEncoding
\makeatletter
\def\ps@headings{
\def\@oddhead{\mbox{}\scriptsize\rightmark \hfil \thepage}%
\def\@evenhead{\scriptsize\thepage \hfil \leftmark\mbox{}}%
\def\@oddfoot{}
\def\@evenfoot{}}
\makeatother
\usepackage{subfigure}
\usepackage{colortbl}
\usepackage{bm}
\usepackage{mathrsfs}
\usepackage{amsmath}
\usepackage{amssymb}
\usepackage{multirow}

\usepackage{graphicx}  
\usepackage{url}       

\usepackage{amsmath}   
\usepackage{cite}

\usepackage{amsfonts,amssymb}

\usepackage{stfloats}
\usepackage{indentfirst}

\usepackage{cases}
\usepackage{algorithm}
\usepackage{multirow}
\usepackage{algorithmic}
\usepackage{stfloats}
\usepackage{epstopdf}

\usepackage{indentfirst}
\makeatletter

\newcommand{\Rmnum}[1]{\expandafter\@slowromancap\romannumeral #1@}
\makeatother

\newcommand{\ls}[1]
    {\dimen0=\fontdimen6\the\font
     \lineskip=#1\dimen0
     \advance\lineskip.5\fontdimen5\the\font
     \advance\lineskip-\dimen0
     \lineskiplimit=.9\lineskip
     \baselineskip=\lineskip
     \advance\baselineskip\dimen0
     \normallineskip\lineskip
     \normallineskiplimit\lineskiplimit
     \normalbaselineskip\baselineskip
     \ignorespaces
    }

\begin{document}
\title{Trajectory and Power Optimization for\\Multi-UAV Enabled Emergency Wireless\\Communications Networks}
\author{\IEEEauthorblockN{Yixin Zhang and Wenchi Cheng}~\\[0.2cm]

\vspace{-10pt}

\IEEEauthorblockA{State Key Laboratory of Integrated Services Networks, Xidian University, Xi'an, China\\
E-mail: \ \emph{\{yixinzhang@stu.xidian.edu.cn,\ wccheng@xidian.edu.cn\}}}

\vspace{-10pt}

}

\maketitle

\begin{abstract}
Recently, unmanned aerial vehicle (UAV) has attracted much attention due to its flexible deployment and controllable mobility. As the general communication network cannot meet the emergency requirements, in this paper we study the multi-UAV enabled wireless emergency communication system. Our goal is to maximize the capacity with jointly optimizing trajectory and allocating power. To tackle this non-convex optimization problem, we first decompose it into two sub-problems to optimize the trajectory and power allocation, respectively. Then, we propose the successive convex approximation technique and the block coordinate update algorithm to solve the two sub-problems. The approximate optimal solution can be obtained after continuous iterations. Simulation results show that the capacity can be greatly increased using our proposed joint trajectory optimization and power allocation.
\end{abstract}

\vspace{10pt}

\begin{IEEEkeywords}
UAV emergency communication, trajectory optimization, power allocation.
\end{IEEEkeywords}

\section{Introduction}
Wireless communication networks have rapidly evolved over the past few decades. Since terrestrial communication systems are equipped with fixed base stations, when unexpected disasters or big events occur, terrestrial communication systems cannot effectively cope with the suddenly growing communication demands for emergency communications.\par
In recent years, there has been a rising trend of using unmanned aerial vehicle (UAV) as aerial mobile base station to improve the performance of existing terrestrial communication systems, such as increasing capacity and extending coverage \cite{overall1},\cite{overall3}. Compared to the traditional terrestrial communication system, the UAV-enabled system has many advantages. UAV can be deployed quickly to meet the needs of burst communications. In addition, since the UAV flies in the sky, the channel is less affected by the ground building, which ensures a better channel quality. What's more, UAV has controllable mobility, so it can make corresponding position changes according to the real-time communication requirements to achieve larger capacity.\par
To optimize the system performance based on UAV-enabled network, there exists some literature investigating the UAV-enabled network in various aspects. On the one hand, when the UAV acts as an aid to assist ground base station communication, the authors established a two-layer hybrid communication model for UAVs and ground networks \cite{ad1}. The authors of \cite{ad2} analyzed the optimal altitude of the UAV, jointly combining the trajectory and power optimization. On the other hand, as for the UAV independently using for aerial base station \cite{place1} - [10], optimizing the deployment of UAVs to improve the coverage performance was studied in\cite{place1},\cite{place2}. In \cite{twobc} , \cite{yy2} and \cite{multicast}, the authors studied the capacity of two-user and multi-user broadcasting channel under the control of single UAV. In addition, trajectory optimization and power allocation have been used in \cite{multiuav1} to make multi-UAV systems perform better. However, in the existing multi-UAV enabled system, the trajectory is usually designed as closed path, that is, the UAV has to return to the initial position at the end of a communication period and round-trip method limits optimization of the UAV trajectory. Under some scenarios, going across the disaster area is more rational than the round-trip.\par
To solve the above mentioned problem, in this paper we study the multi-UAV enabled emergency wireless communication system with the one-way trajectory. We propose a one-way hover-fly-hover (HFH) trajectory method and combine it with power allocation for system optimization. However, in the case of multi-UAV system, there are co-channel interferences which will make the trajectory and power optimization of each UAV constrained mutually. The coupling effect undoubtedly brings great difficulty and the optimization problem is non-convex. To tackle the above challenges, we decompose the original non-convex problem into two sub-problems, and transform them into two convex sub-problems by applying successive convex approximation technique. Then, we apply coordinate descent update algorithm to perform iterative optimization. After iterating, the optimal trajectory and power allocation can be obtained approximatively and the system capacity can reach the maximum.\par
The rest of this paper is organized as follows. Section II introduces the system model and presents the problem formulation for capacity of a multi-UAV enabled wireless network. Section III introduces the trajectory design and corresponding approach solving the optimization problem. Section IV provides the numerical results. Finally, we conclude this paper in Section V.

\section{System Model and Problem Formulation}
\subsection{System Model}
\begin{figure}[htbp]
\centering\includegraphics[width=3.5in]{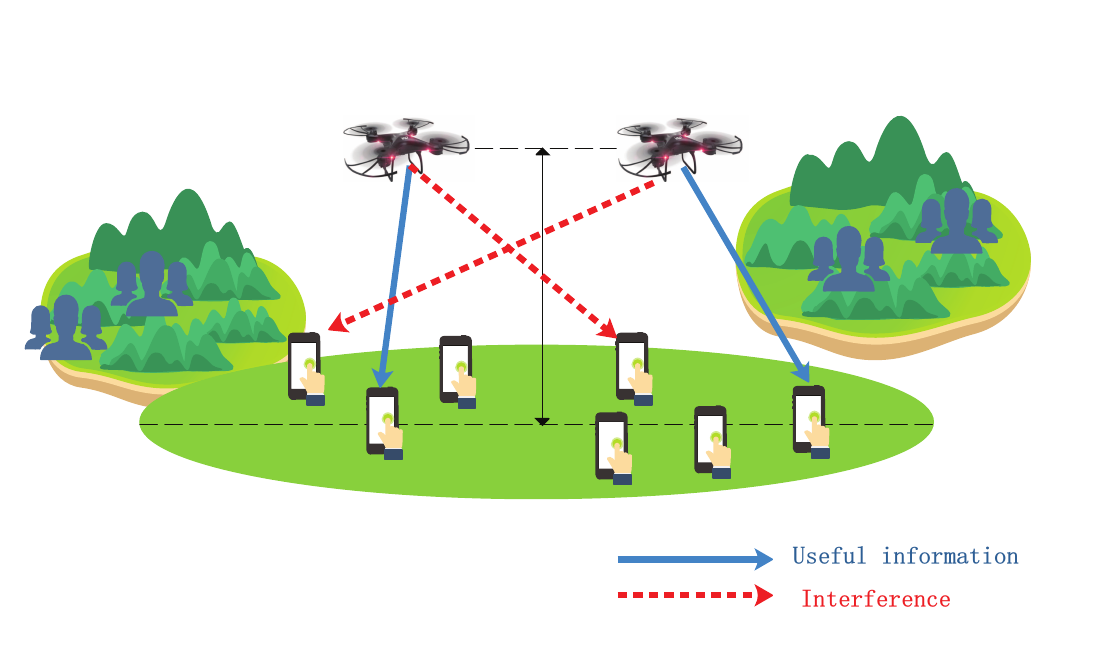}
\caption{The multi-UAV enabled emergency wireless communication system.}\label{fig:1}
\end{figure}
As shown in Fig. \ref{fig:1}, we consider a multi-UAV enabled emergency wireless communication system, in which $M$ UAVs are used to transmit information to $K$ users randomly located on the ground. The sets of UAVs and users are denoted as $\mathcal{M}\triangleq{\{1,...,M\}}$ and $\mathcal{K} \triangleq{\{1,...,K\}}$, respectively. Then, we consider a 3D coordinate system, where the ground user $k\in\mathcal{K}$ has a fixed location $q_k=(x_k,y_k,z_k)$. Suppose that all UAVs have a same communication period at a finite duration $\mathcal{T}\triangleq[0,T]$. During this period, all UAVs are assumed to fly at a fixed altitude $H$(m) above ground and we denote by $q_m(t)=\big(x_m(t),y_m(t),H\big)$ their time-varying locations at any time instant $t\in \mathcal{T}$. Furthermore, we assume that all UAVs share the same frequency band.\par
The mobilities of UAVs are subjected to the following distance constraint represented by

\begin{equation}
||q_m(t)-q_j(t)||\geq d_{\text{min}}, m\neq j, \forall m,
\label{eq:dc}
\end{equation}
where $d_{\text{min}}$ denotes the minimum distance between any pairs of UAVs in order to avoid collision. Besides, we assume that the Doppler effect caused by the mobility of UAV can be compensated at the receiver. Therefore, the channel power gain from the $m$-th UAV to the $k$-th user satisfies the free-space path-loss model, which can be expressed as follows:
\begin{equation}
\begin{split}
h_{m,k}(q_m(t))&=\rho_0 d^{-2}_{m,k}(t)\\
&=\frac{\rho_0}{||q_m(t)-q_k||^2}\\
&=\frac{\rho_0}{|x_m(t)-x_k|^2+|y_m(t)-y_k|^2+|H-z_k|^2},
\label{eq:h}
\end{split}
\end{equation}
where $d_{m,k}(t)$ denotes the distance from the $m$-th UAV to the $k$-th user and $\rho_0$ denotes the channel power gain at a reference distance of $1$ meter. Let $p_m(t)$ denote the downlink transmit power of the $m$-th UAV. We assume that UAVs are subject to a maximum transmit power constraint $P_{\text{max}}$ at any time instant $t\in \mathcal{T}$, which can be expressed as follows:
\begin{equation}
0\leq p_m(t)\leq P_{\text{max}}, \forall m.
\label{eq:pc}
\end{equation}

Therefore, the corresponding received signal-to-interference-plus-noise ratio (SINR) at the $k$-th user can be expressed as follows:
\begin{equation}
\gamma_{m,k}(t)=\frac{p_m(t)h_{m,k}(t)}{\sum_{j=1,j\neq m}^{M}p_j(t)h_{j,k}(t)+{\sigma}^2},
\label{eq:SINR}
\end{equation}
where ${\sigma}^2$ is the noise power. In the denominator of (\ref{eq:SINR}), $\sum_{j=1,j\neq m}^{M}p_j(t)h_{j,k}(t)$ represents the co-channel interferences caused by the other UAVs (except the $m$-th) to the $k$-th user. Thus, the capacity of the $k$-th user over the communication period $\mathcal{T}$ can be expressed as follows:
\begin{equation}
\overline{R}_k[{q_m(t),p_m(t)}]=\frac{1}{T}\int_{0}^{T}R_{k}[{q_m(t),p_m(t)}]dt,
\end{equation}
where
\begin{equation}
R_{k}[{q_m(t),p_m(t)}]=\log_2{\big(1+\gamma_{m,k}(t)\big)}.
\end{equation}

\subsection{Problem Formulation}
Based on the above discussion, we take both the trajectory optimization and the power allocation into consideration and optimize them jointly. Our goal is to achieve maximum capacity, that is, maximize the minimum value of $\overline{R}_k[{q_m(t),p_m(t)}]$. So we introduce an auxiliary variable $\eta$ to make the expression simpler. The optimization problem is formulated as follows:
\begin{align}
\text{(P1):}&\hspace{0.5cm}\underset{\{q_m(t),p_m(t),\eta\}}{\text{max}}\ \ \ \  \eta ,& \nonumber
\end{align}
subject to $\frac{1}{T} \int_{0}^{T}R_{k}[{q_m(t),p_m(t)}]dt \geq \eta$, Eq. (\ref{eq:dc}), and Eq. (\ref{eq:pc}).

\section{Trajectory Design and Optimization Approach}
In this section, we first propose a one-way HFH trajectory optimization method which is suitable for multi-UAV enabled system. Then, we consider a selection method to divide the ground users into $M$ groups and assign each user group to the appropriate UAV for emergency wireless communication. After the UAV-user pairs are determined, we optimize the trajectory and power allocation jointly. Since the joint optimization problem is non-convex which can not be solved by the traditional convex optimization method, we transform the problem into two convex sub-problems to optimize trajectory and power separately by applying successive convex approximation technique. Then, we use the block coordinate update algorithm to optimize them jointly by updating iteratively. Finally, the maximum capacity can be obtained.
\subsection{One-way HFH Trajectory Design}
Under some emergency scenarios, going across the disaster area is more rational than the round-trip. So we propose a trajectory design which UAV does not have to return to the initial location. We extend the HFH trajectory design \cite{twobc} into multi-UAV and multi-user system. The basic idea of the hover-fly-hover (HFH) trajectory design is that UAV hovers at a pair of optimal initial and final locations and flies between them unidirectionally. In multi-UAV enabled system with the effects of interferences, within one period, the UAV first calculates the trajectory based on its initial position and the location of users, then, the UAV flies or hovers according to the calculated trajectory. When this period ends, the UAV calculates a new trajectory based on the new initial position and the location of users and flies or hovers again. Repeat this process continuously, the trajectory of the UAV will be many continuous line segments and UAV does not have to return to the initial location, so it is called one-way HFH trajectory.\par
The initial and final locations of the $m$-th UAV are represented as $q_{Im}$ and $q_{Fm}$, and the hovering time at the initial and final locations are represented as $t_{Im}$ and $t_{Fm}$, respectively. The expression of the one-way HFH trajectory $\mathbf{Q}_{\text{OHFH}}$ in one communication period ${\mathcal{T}}$ is expressed as follows:
\begin{flalign}
q_m(t)&=
\begin{cases}
q_{Im}\\
q_{Im}+(t-t_{Im})V\\
q_{Fm}
\end{cases}
\nonumber\\&=
\begin{cases}
(x_{Im},y_{Im},H), &t\in T_1, \\
\big(x_{Im}+(t-t_{Im})V\cos\theta,&\\
 \hspace{1cm}y_{Im}+(t-t_{Im})V\sin\theta ,H \big), &t\in T_2,\\
(x_{Fm},y_{Fm},H), &t\in T_3,
\end{cases}
\end{flalign}
where $T_1=[0,t_{Im})$, $T_2=[t_{Im},T-t_{Fm})$, $T_3=[T-t_{Fm},T]$, $t_{Im}+\frac{||q_{Fm}-q_{Im}||}{V}+t_{Fm}=T$ and $\theta$ is the angle between the speed $V$ and the horizontal direction, with $\theta=\arctan{\frac{y_{Fm}-y_{Im}}{x_{Fm}-x_{Im}}}$.\par
By the above formula, suppose that we know the initial position of the $m$-th UAV and the locations of the ground users, the trajectory can be completely determined if we know the final position $q_{Fm}$ and the initial hovering time $t_{Im}$. So our goal is to find the optimal final position $q_{Fm}$ and the initial hovering time $t_{Im}$ which can make the system reaches the maximum capacity.

\subsection{Selection of UAV-User Pairs}
In reality, a ground user is only served by one UAV in one communication period, and thus we should first select the appropriate UAV-user pair.\par
Assume that the fixed locations of the ground users and the initial positions of the UAVs are known, then, we can calculate the distance between each user to each UAV. According to Eqs. (\ref{eq:h}) and (\ref{eq:SINR}), it can be obtained that the user should select the UAV which is closest to himself to communicate, so that not only the transmission power gain can be increased, but also the co-channel interferences can be appropriately reduced. The UAV-user pairs selection method can be expressed as follows:
\begin{eqnarray}
\hspace{-3cm}\!\!\!\text{(P2):}\!\!\!\!\!\! && \underset{\{(m-k)\ pair\}}{\text{arg min}}\ \ \  d_{m,k} \nonumber
\end{eqnarray}
\begin{equation}
\begin{split}
\hspace{-0.2cm}\ \ \ \mathrm{s.t.\ \ \ \ \ \ }    d^2_{m,k}=|x_{Im}-x_k|^2+|y_{Im}&-y_k|^2
\\ &+|H-z_k|^2.
\end{split}
\end{equation}
This problem is a linear problem and it can be solved easily. Then, we use these UAV-user pairs to solve the following trajectory and power optimization problem.

\subsection{Overall Method}
Based on the proposed one-way HFH trajectory, the overall optimization problem can be expressed as
\begin{eqnarray}
\hspace{-3cm}\!\!\!\text{(P3):}\!\!\!\!\!\! && \max_{\{q_m(t),p_m(t),\eta\}}\ \ \  \eta \nonumber
\end{eqnarray}
\begin{flalign}
\hspace{0.3cm}\ \ \ \mathrm{s.t.\ \ \ \ \ \ }  &  \frac{1}{T}\int_0^T {R_k[(q_m(t),p_m(t)]}dt\geq {\eta},&\\
& 0\leq p_m(t)\leq P_{\text{max}},\ \forall m,&\\
& {q}_m(t)\in\mathbf{Q}_{\text{OHFH}},\ \forall m,&\\
&d^2_{\text{min}}\leq ||q_m(t)-q_j(t)||^2, \ j\neq m, \ \forall j,m.&
\end{flalign}
(P3) is a non-convex problem, so we decompose it into two sub-problems to optimize trajectory and allocate power separately by applying successive convex approximation technique \cite{sco1}. Then, we use coordinate descent update algorithm \cite{al} to perform iterative optimization. After iterating $l$ times, the optimal solution can be obtained. The block coordinate update algorithm is shown in Table \ref{t:a}. The trajectory update algorithm is divided into three steps and the values of $x_{Fm}$, $y_{Fm}$ and $t_{Im}$ are updated in turn, which be expressed as $\mathbf q_{m1}$, $\mathbf q_{m2}$ and $\mathbf q_{m}$.

\newcommand{\tabincell}[2]{\begin{tabular}{@{}#1@{}}#2\end{tabular}}
\begin{table}[htbp]
\caption{\label{t:a}Algorithm for Solving Problem (P3).}
\begin{tabular}{l l}
\hline
\multicolumn{2}{l}{Algorithm 1: Block Coordinate Update Algorithm for Trajectory }\\
\multicolumn{2}{l}{ \hspace{1.6cm}and Power Allocation Optimization}\\
\hline
1.&\tabincell{l} {Let iteration index $l=0$, initialize the UAV’s trajectory as\\ $\mathbf q_m^{[l]}$ \{$x_{Fm}^{[l]}, y_{Fm}^{[l]},t_{Im}^{[l]}$\} and the UAV’s power allocation as \\ $\mathbf p_m^{[l]}$;}\\
2.&\textbf{repeat};\\
3.&\tabincell{l} {For given power allocation $\mathbf p_m^{[l]}$, update the UAVs trajectory as \\ $\mathbf q_{m1}^{[l+1]}$ \{$x_{Fm}^{[l+1]}, y_{Fm}^{[l]}, t_{Im}^{[l]}$\};}\\
4.&\tabincell{l}{For given power allocation $\mathbf p_m^{[l]}$ and $\mathbf q_{m1}^{[l+1]}$ \{$x_{Fm}^{[l+1]}, y_{Fm}^{[l]}, t_{Im}^{[l]}$\},\\ update the UAVs trajectory as $\mathbf q_{m2}^{[l+1]}$  \{$x_{Fm}^{[l+1]}, y_{Fm}^{[l+1]}, t_{Im}^{[l]}$\};}\\
5.&\tabincell{l}{For given power allocation $\mathbf p_m^{[l]}$ and $\mathbf q_{m2}^{[l+1]}$\{$x_{Fm}^{[l+1]}, y_{Fm}^{[l+1]}, t_{Im}^{[l]}$\},\\ update the UAVs trajectory as $\mathbf q_{m}^{[l+1]}$\{$x_{Fm}^{[l+1]}, y_{Fm}^{[l+1]}, t_{Im}^{[l+1]}$\};}\\
6.&\tabincell{l}{For given trajectory $\mathbf q_m^{[l+1]}$, update the optimal power allocation \\as $\mathbf p_m^{[l+1]}$;}\\
7.&Update $l=l+1$;\\
8.&\textbf{until} the increase of the value of $\eta$ is less than a given threshold.\\
\hline
\end{tabular}
\end{table}

\subsection{Trajectory Optimization with Fixed Power}
We first consider the trajectory optimization when the power allocation is fixed. For any given power allocation $\mathbf{p}_m$, the UAV trajectory optimization of problem (P3) can be optimized by solving the following problem
\begin{eqnarray}
\hspace{-3cm}\!\!\!\text{(P3.1):}\!\!\!\!\!\! && \underset{\{{q_m(t)},{\eta}_t\}}{\text{max}}\ \ \ {\eta}_t \nonumber
\end{eqnarray}
\begin{flalign}
\hspace{0.3cm}\ \ \ \mathrm{s.t.\ \ \ \ \ \ }  &  \frac{1}{T}\int_0^T {R_k[q_m(t)]}dt\geq {\eta_t},\label{eq:n2} \\
& {q}_m(t)\in\mathbf{Q}_{\text{OHFH}}, \ \forall m,&\\
&d^2_{\text{min}}\leq ||q_m(t)-q_j(t)||^2, \ j\neq m,\ \forall j,m.\label{eq:n3}&
\end{flalign}
Since Eqs. (\ref{eq:n2}) and (\ref{eq:n3}) are non-convex constraints, the problem (P3.1) is a non-convex problem and it is difficult to solve. So we first decompose $R_k[q_m(t)]$ in Eq. (\ref{eq:n2}) into two parts as follows and use the successive convex approximation technique to solve the problem (P3.1) \cite{multiuav1}, \cite{sco1}.
\begin{equation}
\begin{split}
R_k[q_m(t)]&=\hat{R}_k[q_m(t)]\\&-\log_2\Big(\sum_{j=1, j\neq m}^{M}\frac{p_j(t){\rho}_0}{||q_j(t)-q_k||^2}+{\sigma}^2\Big),
\end{split}
\end{equation}
where
\begin{equation}
\hat{R}_k[q_m(t)]=\log_2\Big(\sum_{j=1}^{M}\frac{p_j(t){\rho}_0}{||q_j(t)-q_k||^2}+{\sigma}^2\Big).
\end{equation}\par
We use its first-order Taylor expansion \cite{taylor} to characterize the lower bound of $\hat{R}_k[q_m(t)]$ as follows:
\begin{equation}
\begin{split}
\hat{R}_k^{lb}[q_m(t)]\triangleq& \log_2\Big(\sum_{j=1}^{M} \frac{p_j(t){\rho}_0}{||q_j^l(t)-q_k||^2}+{\sigma}^2\Big)\\ &+\sum_{j=1}^{M}A_{j,k}^l(t)\Big(||q_j(t)-q_k||^2-||q_j^l(t)-q_k||^2\Big),
\end{split}
\end{equation}
where
\begin{equation}
A_{j,k}^l(t)=\frac{\frac{-p_j(t){\rho}_0}{\big(||q_j^l(t)-q_k||^2\big)^2}}{(\sum_{i=1}^{M}\frac{p_i(t){\rho}_0}{||q_i^l(t)-q_k||^2}+{\sigma}^2)\ln 2},
\end{equation}
and $q_j^l(t)$ represents the value of $q_j(t)$ in the $l$-th iteration.\par
After the above equation transformation, the constraint Eq. (\ref{eq:n2}) is still non-convex. So we define $\lambda(t)=||q_j(t)-q_k||^2$ and $\xi(t)=||q_m(t)-q_j(t)||^2$ for notational convenience and we have
\begin{flalign}
\hspace{-1cm} \lambda(t)&\geq ||q_j^l(t)-q_k||^2+2\big(q_j^l(t)-q_k\big)^\mathrm{T}\times \big(q_j(t)-q_j^l(t)\big) \nonumber \\
 &\triangleq \lambda^{lb}(t),
\end{flalign}
and
\begin{flalign}
\xi(t)&\geq -||q_m^l(t)-q_j^l(t)||^2+2\big(q_m^l(t)-q_j^l(t)\big)^\mathrm{T} \nonumber \\ &\hspace{4cm} \times \big(q_m(t)-q_j(t)\big)\nonumber  \\
&\triangleq \xi^{lb}(t).
\end{flalign}\par
Based on the above discussion, we can transform the trajectory optimization problem into the following convex optimization problem (P3.2).
\begin{eqnarray}
\hspace{-4cm}\!\!\!\text{(P3.2):}\!\!\!\!\!\! && \max_{\{{q_m(t)},{\eta}_t^l\}}\ \ \  {\eta}_t^l \nonumber
\end{eqnarray}
\begin{flalign}
\ \ \ \mathrm{s.t.\ \ \ \ \ \ }  &  \frac{1}{T}\int_0^T \Big(\hat{R}_{k}^{lb}[q_m(t)] -\log_2\big(\sum_{j=1,j\neq m}^{M} &\nonumber\\
&\hspace{1.7cm}\frac{p_j(t){\rho}_0}{\lambda^{lb}(t)}+{\sigma}^2\big)\Big)dt\geq \eta_t^l,& \label{eq:c2}\\
& {q}_m(t)\in\mathbf{Q}_{\text{OHFH}},\ \forall m, &\\
&d_{\text{min}}^2\leq \xi^{lb}(t).&\label{eq:c3}
\end{flalign}
At this point, the constraints Eqs. (\ref{eq:c2}) and (\ref{eq:c3}) are convex, so we can obtain the optimal solution by CVX \cite{cvx}. And it is worth mentioning that the obtained result according to problem (P3.2) is lower than problem (P3), but through continuous iteration by block coordinate updating, the result is very close to the optimal solution.

\subsection{Power Optimization with Fixed Trajectory}
Then, we consider the power optimization problem when the trajectory is fixed. For any given UAV trajectory $\mathbf{q}_m$, the UAV transmit power allocation of problem (P3) can be optimized by solving the following problem
\begin{eqnarray}
\hspace{-3cm}\!\!\!\text{(P3.3):}\!\!\!\!\!\! && \underset{\{{p_m(t)},{\eta}_p\}}{\text{max}}\ \ \ {\eta}_p \nonumber
\end{eqnarray}
\begin{flalign}
\hspace{0.3cm}\ \ \ \mathrm{s.t.\ \ \ \ \ \ }  &  \frac{1}{T}\int_0^T {R_k[p_m(t)]}dt\geq \eta_p, &\label{eq:n1}\\
&0\leq p_m(t)\leq P_{\text{max}}, \ \forall m.
\end{flalign}
Since Eq. (\ref{eq:n1}) is a non-convex constraint, the problem (P3.3) is a non-convex problem and it is difficult to solve. Same as the case of (P3.1), we decompose $R_k[p_m(t)]$ in Eq. (\ref{eq:n1}) into two parts as follows:
\begin{flalign}
{R_k[p_m(t)]}=\log_2\Big(\sum_{j=1}^{M}p_j(t)h_{j,k}(t)+{\sigma}^2 \Big)-\check{R}_k[p_m(t)],
\end{flalign}
where
\begin{equation}
\check{R}_k[p_m(t)]=\log_2\Big({\sum_{j=1,j\neq m}^{M}p_j(t)h_{j,k}(t)+{\sigma}^2}\Big).
\end{equation}\par
Then, we continue to use the first-order Taylor expansion to find the upper bound of $\check{R}_k[p_m(t)]$ as follows:
\begin{equation}
\begin{split}
\check{R}_k^{ub}[p_m(t)]\triangleq &\log_2{\Big(\sum_{j=1,j\neq m}^{M}p_j^l(t)h_{j,k}(t)+{\sigma}^2 \Big)}  \\
&+\sum_{j=1,j\neq m}^{M}B_{j,k}(t)\Big(p_j(t)-p_j^l(t)\Big),
\end{split}
\end{equation}
where
\begin{equation}
B_{j,k}^l(t)=\frac{h_{j,k}(t)}{\big(\sum_{i=1,i\neq m}^{M}p_i^l(t)h_{i,k}(t)+{\sigma}^2 \big) \ln 2},
\end{equation}
and $p_j^l(t)$ represents the value of $p_j(t)$ in the $l$-th iteration. Therefore, (P3.3) can be transformed as follows
\begin{eqnarray}
\hspace{-3cm}\!\!\!\text{(P3.4):}\!\!\!\!\!\! && \underset{\{{p_m(t)},{\eta}_p^l\}}{\text{max}}\ \ \ {\eta}_p^l \nonumber
\end{eqnarray}
\begin{flalign}
\hspace{0.3cm}\ \ \ \mathrm{s.t.\ \ \ \ \ \ }  &  \frac{1}{T}\int_{0}^{T} \Big{(}\big(\log_{2}(\sum_{j=1}^{M}p_j(t)h_{j,k}(t) &\nonumber\\
& \hspace{1.4cm}+{\sigma}^2\big)-\check{R}_k^{ub}[p_m(t)]\Big{)}dt\geq {\eta}_p^l, &\label{eq:c1}\\
&0\leq p_m(t)\leq P_{\text{max}},\ \forall m.&
\end{flalign}
It can be seen that the the problem (P3.4) is a convex optimization problem, which can be solved by standard convex optimization method \cite{ld1},\cite{yy1}.

\section{Simulation Results}
In this section, we represent the simulation results to validate the validity of the proposed optimization method. We first consider a communication system with $M = 1, 2, 3$ UAVs and $K = 8$ users, where $8$ users are randomly distributed in a square area of $1000$ m by $1000$ m  on the ground. Assume that all the UAVs fly at a fixed altitude $H=100$ m and the minimum distance $d_{\text{min}}$ between any UAV pairs is $50$ m. The channel power gain $\rho_0$ is assumed to be $-60$ dB at the reference distance $d_0 = 1$ m. The noise power at the receiver is set as $\sigma^2 = -100$ dBm. We assume the UAV flight speed is fixed as $V = 10$ m/s, and the communication period is set as $T = 100$ s. Besides, the maximum power constraint is set as $P_{\text{max}} = -30$ dBm. The receiver sensitivity is assumed to be $-100$ dBm and the used frequency is $100$ MHz. 
\begin{figure}[htbp]
\centering\includegraphics[width=3.5in]{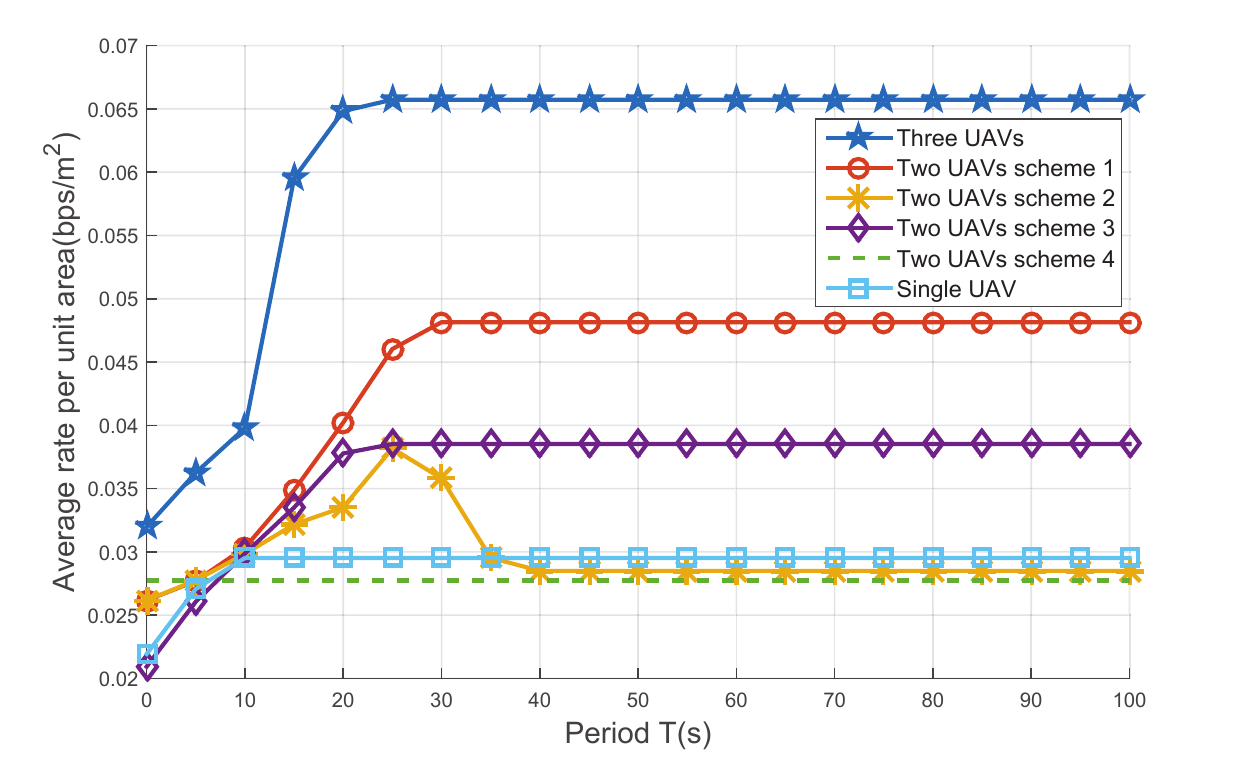}
\caption{Average rate of $8$ users in different multi-UAV enabled communication system.}\label{fig:ar1}
\end{figure}\par
Figure \ref{fig:ar1} represents the average rate of $8$ users versus the communication period $\mathcal{T}$ in different multi-UAV enabled communication systems. First, we can observe that the average rate per unit area increases with the number of UAVs increasing. Then, we compare the average rate of four schemes in the case of two UAVs existed in system. Scheme 1 is the proposed method with joint trajectory and power allocation optimization, and the average rate can reach $0.0442\ \text{bps/m}^2$ during the period $\mathcal{T}$. Scheme 2 is the power allocation method with fixed trajectory and the scheme 3 is the trajectory method with fixed power allocation. Their average rates are $0.0298\ \text{bps/m}^2$ and $0.0367\ \text{bps/m}^2$, respectively. Scheme 4 is assumed that two UAVs are all fixed and the average rate in this case is the lowest. Through the joint optimization of the trajectory and power, the system can achieve a larger rate.
\begin{figure}[htbp]
\centering\includegraphics[width=3.2in]{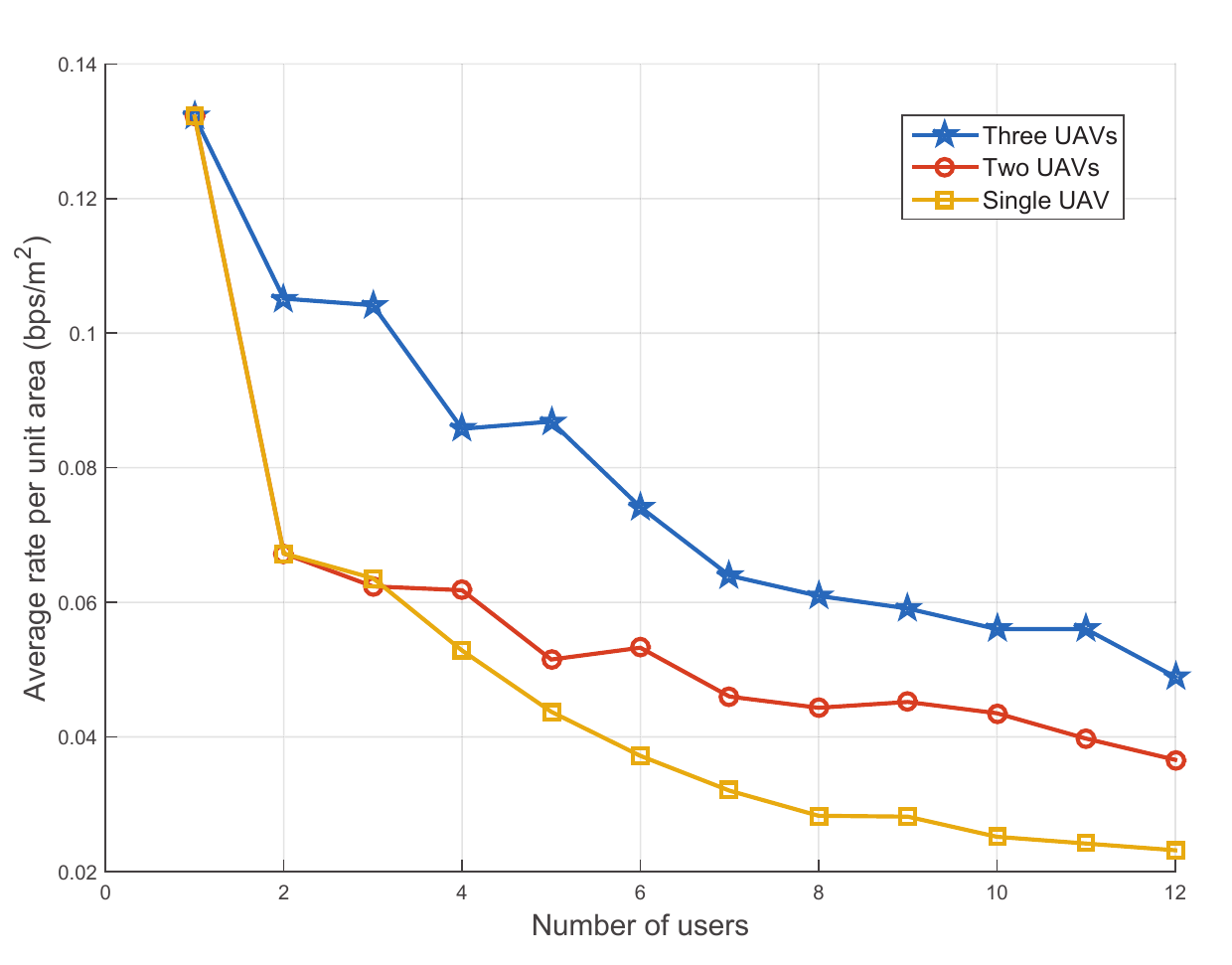}
\caption{Comparison of average rate between single UAV system and multi-UAV system.}\label{fig:ar2}
\end{figure}\par
Figure \ref{fig:ar2} compares the average rate changes under different systems with one UAV, two UAVs and three UAVs as number of users increasing from $1$ to $12$. As illustrated in Fig. \ref{fig:ar2}, we can observe that the average rate per unit area of the system decreases as the number of users increases. But with the number of UAVs increasing, the average rate per unit area can maintain at a larger value, which can better guarantee the emergency communication performance.

\section{Conclusions}
In this paper, we studied the multi-UAV enabled wireless network system for emergency communication, which can be used in disaster or emergency scenarios. In order to optimize the performance of this network system, we proposed a one-way HFH trajectory design and optimize it with power allocation jointly. Since this optimization problem is non-convex, we decompose this problem into two sub-problems and use the successive convex approximation technique and the block coordinate update algorithm to solve it. After iterating, the optimal solution can be approximately obtained. The obtained numerical results validate that our proposed joint trajectory and power allocation optimization scheme significantly increases the rate for the multi-UAV enabled communication systems.

\bibliographystyle{IEEEtran}
\bibliography{References}

\begin{thebibliography}{10}
\providecommand{\url}[1]{#1}
\csname url@samestyle\endcsname
\providecommand{\newblock}{\relax}
\providecommand{\bibinfo}[2]{#2}
\providecommand{\BIBentrySTDinterwordspacing}{\spaceskip=0pt\relax}
\providecommand{\BIBentryALTinterwordstretchfactor}{4}
\providecommand{\BIBentryALTinterwordspacing}{\spaceskip=\fontdimen2\font plus
\BIBentryALTinterwordstretchfactor\fontdimen3\font minus
  \fontdimen4\font\relax}
\providecommand{\BIBforeignlanguage}[2]{{%
\expandafter\ifx\csname l@#1\endcsname\relax
\typeout{** WARNING: IEEEtran.bst: No hyphenation pattern has been}%
\typeout{** loaded for the language `#1'. Using the pattern for}%
\typeout{** the default language instead.}%
\else
\language=\csname l@#1\endcsname
\fi
#2}}
\providecommand{\BIBdecl}{\relax}
\BIBdecl

\bibitem{overall1}
Y.~{Zeng}, R.~{Zhang}, and T.~J. {Lim}, ``Wireless communications with unmanned
  aerial vehicles: opportunities and challenges,'' \emph{IEEE Commun. Mag.},
  vol.~54, no.~5, pp. 36--42, May 2016.

\bibitem{overall3}
Y.~{Zhou}, N.~{Cheng}, N.~{Lu}, and X.~S. {Shen}, ``Multi-uav-aided networks:
  Aerial-ground cooperative vehicular networking architecture,'' \emph{IEEE
  Veh. Technol. Mag.}, vol.~10, no.~4, pp. 36--44, Dec 2015.

\bibitem{ad1}
J.~{Lyu}, Y.~{Zeng}, and R.~{Zhang}, ``Spectrum sharing and cyclical multiple
  access in uav-aided cellular offloading,'' in \emph{IEEE GLOBECOM}, Dec 2017,
  pp. 1--6.

\bibitem{ad2}
M.~{Mozaffari}, W.~{Saad}, M.~{Bennis}, and M.~{Debbah}, ``Unmanned aerial
  vehicle with underlaid device-to-device communications: Performance and
  tradeoffs,'' \emph{IEEE Trans. Wireless Commun.}, vol.~15, no.~6, pp.
  3949--3963, June 2016.

\bibitem{place1}
J.~{Lyu}, Y.~{Zeng}, R.~{Zhang}, and T.~J. {Lim}, ``Placement optimization of
  uav-mounted mobile base stations,'' \emph{IEEE Commun. Lett.}, vol.~21,
  no.~3, pp. 604--607, March 2017.

\bibitem{place2}
R.~I. {Bor-Yaliniz}, A.~{El-Keyi}, and H.~{Yanikomeroglu}, ``Efficient 3-d
  placement of an aerial base station in next generation cellular networks,''
  in \emph{IEEE ICC}, May 2016, pp. 1--5.

\bibitem{twobc}
Q.~{Wu}, J.~{Xu}, and R.~{Zhang}, ``Capacity characterization of uav-enabled
  two-user broadcast channel,'' \emph{IEEE J. Sel. Areas Commun.}, vol.~36,
  no.~9, pp. 1955--1971, Sep. 2018.

\bibitem{yy2}
W.~{Cheng}, X.~{Zhang}, and H.~{Zhang}, ``Full-duplex spectrum-sensing and
  mac-protocol for multichannel nontime-slotted cognitive radio networks,''
  \emph{IEEE J. Sel. Areas Commun.}, vol.~33, no.~5, pp. 820--831, May 2015.

\bibitem{multicast}
Y.~{Wu}, J.~{Xu}, L.~{Qiu}, and R.~{Zhang}, ``Capacity of uav-enabled multicast
  channel: Joint trajectory design and power allocation,'' in \emph{IEEE ICC
  Workshop}, May 2018, pp. 1--7.

\bibitem{multiuav1}
Q.~{Wu}, Y.~{Zeng}, and R.~{Zhang}, ``Joint trajectory and communication design
  for multi-uav enabled wireless networks,'' \emph{IEEE Trans. Wireless
  Commun.}, vol.~17, no.~3, pp. 2109--2121, March 2018.

\bibitem{sco1}
Y.~{Zeng}, R.~{Zhang}, and T.~J. {Lim}, ``Throughput maximization for
  uav-enabled mobile relaying systems,'' \emph{IEEE Trans. Wireless Commun.},
  vol.~64, no.~12, pp. 4983--4996, Dec 2016.

\bibitem{al}
Y.~Xu and W.~Yin, ``A block coordinate descent method for multi-convex
  optimization with applications to nonnegative tensor factorization and
  completion,'' \emph{Siam Journal on Imaging Sciences}, vol.~6, no.~3, pp.
  1758--1789, 2015.

\bibitem{taylor}
T.~{Wang} and L.~{Vandendorpe}, ``Successive convex approximation based methods
  for dynamic spectrum management,'' in \emph{IEEE ICC}, June 2012, pp.
  4061--4065.

\bibitem{cvx}
\BIBentryALTinterwordspacing
M.~Grant and S.~Boyd., \emph{{CVX: MATLAB Software for Disciplined Convex
  Programming, Version 2.1}}, 2nd~ed., 2016. [Online]. Available:
  \url{http://cvxr.com/cvx/}
\BIBentrySTDinterwordspacing

\bibitem{ld1}
S.~Boyd and L.~Vandenberghe, \emph{Convex Optimization}.\hskip 1em plus 0.5em
  minus 0.4em\relax Cambridge, U.K.: Cambridge Univ. Press, 2004.

\bibitem{yy1}
W.~{Cheng}, X.~{Zhang}, and H.~{Zhang}, ``Optimal dynamic power control for
  full-duplex bidirectional-channel based wireless networks,'' in \emph{IEEE
  INFOCOM}, April 2013, pp. 3120--3128.

\end{thebibliography}

\end{document}